\documentclass[useAMS,usenatbib,intlimits,centertags]{mn2e}
\usepackage{graphicx}
\usepackage{amsmath}
\usepackage{booktabs}

\allowdisplaybreaks
\newcommand{\rs}{r_\text{s}}
\renewcommand{\d}{\text{d}}

\begin{document}

 \title{Uniformly Rotating Polytropic Rings in Newtonian Gravity}

 \author[D.\ Petroff \& S.\ Horatschek]
 {David Petroff\thanks{E-mail: {\tt D.Petroff@tpi.uni-jena.de} (DP);\newline {\tt S.Horatschek@tpi.uni-jena.de} (SH)} and Stefan Horatschek\addtocounter{footnote}{-1}\footnotemark\\
 Theoretisch-Physikalisches Institut, University of Jena, Max-Wien-Platz 1, 07743 Jena, Germany}
 \date{\today}

 \pagerange{\pageref{firstpage}--\pageref{lastpage}} \pubyear{2008}

 \maketitle

 \label{firstpage}

 \begin{abstract}
  An iterative method is presented for solving the problem of a uniformly rotating, self-gravitating
  ring without a central body in Newtonian gravity by expanding about the thin ring limit. Using this method, a simple formula
  relating mass to the integrated pressure is derived to the leading order for a general equation of state.
  For polytropes with the index $n=1$, analytic coefficients of the iterative approach are determined up to
  the third order. Analogous coefficients are computed numerically for other polytropes. Our solutions are compared with those
  generated by highly accurate numerical methods to test their accuracy.
 \end{abstract}


 \section{Introduction}

 Motivated in part by the rings of Saturn, \citet{Kowalewsky85}, \citet{Poincare85} and \citet{Dyson92, Dyson93}
 studied, amongst other things, the problem of an axially symmetric, homogeneous fluid ring
 in equilibrium by expanding it about the thin ring limit. In particular, Dyson provided a solution to fourth order in the
 parameter $\sigma=a/b$, where $a$ provides a measure for the radius of the cross-section of the ring and $b$ the
 distance of the cross-section's centre of mass from the axis of rotation. An important step toward understanding rings with other equations of state was taken by
\citet{Ostriker64,Ostriker64b,Ostriker65}, who studied polytropic rings to first order in $\sigma$
and found a complete solution to this order for an isothermal limit.

 Numerical methods were developed to study such rings
 and their connection to the Maclaurin spheroids \citep*{Wong74,ES81,EH85,AKM03}.
 With numerical methods, it was also possible to treat the problem of non-homogeneous rings and even within the framework of
 General Relativity \citep*{Hachisu86,AKM03c,FHA05}.

 Through the use of computer algebra, we
 were able to extend Dyson's basic idea and determine the solution to the problem of the homogeneous ring up
 to the order $\sigma^{20}$ \citep{HP08}.
 In this paper, we present an iterative method for performing a similar expansion about the
 thin ring limit for arbitrary equations of state and a number of general results are derived
 confirming and generalizing work that had already been published by \citet{Ostriker64b}.
 The application to polytropes is considered and ordinary differential equations (ODEs) are found
 that allow for the determination of the mass density. A closed-form solution can only be
 found if the value of the polytropic index is $n=1$, and such rings are considered to the order
 $\sigma^3$. For other polytropic indices, the ODEs are solved numerically
 so that results from the approximate scheme can be compared to highly accurate numerical results
 for a variety of equations of state. The numerical solutions considered here are taken from
 a multi-domain spectral program, much like the one described in \citet*{AKM03b}, but tailored
 to Newtonian bodies with toroidal topologies (see \citealt{AP05} for more information). The solutions
 obtained by these numerical methods are extremely accurate and thus provide us with a means
 of testing the accuracy of the approximate method.

 \section{Approximation Scheme}\label{approx}
 Numerical results indicate that, independent of the equation of state, the shape of the cross-section of a uniformly rotating
 ring tends to that of a circle in the thin ring limit, i.e.\ the limit in which the ratio of the inner radius
 $\varrho_\text i$ to the outer one $\varrho_\text o$ tends to 1. This suggests that, as in the homogeneous
 case, a Fourier expansion of the quantities involved will yield coefficients related to the thinness of
 the ring. This last statement can be made more precise when the explicit approximation scheme is presented.

\begin{figure}
  \centerline{\includegraphics{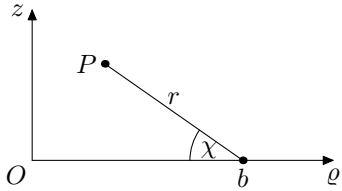}}
  \caption{A sketch providing the meaning of the coordinates $(r,\chi)$.\label{coordinates}}
 \end{figure}

 Let us begin by introducing the constant $b$ and polar-like coordinates $(r,\chi,\varphi)$ related to the cylindrical coordinates
 $(\varrho,z,\varphi)$ by
 \begin{equation}
  \varrho = b - r \cos \chi, \qquad z=r\sin \chi, \qquad \varphi=\varphi.
 \end{equation}
 For a given value of $\varphi$, constant values of the coordinate $r$ are circles centred about
 $(\varrho=b,z=0)$ and $\chi$ measures the angle
 along any such circles. Fig.~\ref{coordinates} provides an illustration of the coordinates.
 The surface of the ring in cross-section can be described by a function $r=\rs(\chi)$, which
 we expand along with the mass density $\mu$ and squared angular velocity $\Omega^2$,
 \begin{align}
  \mu(r,\chi)   &= \mu_\text c\left(\sum_{i=0}^q\sum_{k=0}^i \mu_{ik}(y) \cos(k\chi) \sigma^i + o(\sigma^q)\right), \label{mu_expansion}\\
  \Omega^2      &= \pi G\mu_\text c \left(\sum_{i=0}^{q+1} \Omega_{i} \sigma^i+ o(\sigma^{q+1})\right) \intertext{and}
  \rs(\chi)     &= a\left(1 + \sum_{i=1}^q\sum_{k=0}^i   \beta_{ik}\cos(k\chi) \sigma^i+ o(\sigma^q)\right),\label{rs}
 \end{align}
 where we have introduced  the dimensionless radius
 \begin{align}
  y := \frac{r}{a}
 \end{align}
 and the parameter
 \begin{equation}
  \sigma := \frac{a}{b},
 \end{equation}
 which tends to 0 in the thin ring limit%
 \footnote{Note that the expansion for $\rs$ contains terms with $k=0$ in contrast to the analogous
  expression for homogeneous rings \citep{HP08}, where the summation index $k$ runs only from 1 to $i$.
  The reason behind this will be discussed shortly.}.
 The quantity $\mu_\text c$ is chosen to be the mass density
 at the point $r=0$ and does not represent the density's maximal value, although it will not differ
 significantly from it in general. The absence of sine terms in the Fourier expansions is as
 a result of the symmetry with respect to the equatorial plane, which is known to hold for
 stationary solutions \citep[see][]{Lichtenstein33}.

 In this section, we present a method for finding $\mu_{qk}$, $\Omega_{q+1}$ and $\beta_{qk}$ given that
 the previous terms in $\sigma^i$ are known.

 The idea used in Dyson's approximation scheme for homogeneous rings makes use of the Poisson integral
 to determine the gravitational potential in terms of the (still unknown) function $\rs$ along
 the axis of rotation \citep{Dyson92}. This is only possible since the mass density is completely determined
 for homogeneous matter
 once the shape of the ring is given. In general, however, it is necessary first to determine $\mu$ to
 the desired order before being able to perform the integral. Here we obtain ordinary, second order
 differential equations for $\mu_{qk}(r)$ by applying the Laplace operator to the integrated Euler equation
 \begin{align}
  & U + \int_0^p\frac{\d p'}{\mu(p')} - \frac{1}{2}\Omega^2\varrho^2 = V_0,  \label{int_Euler} \\
  & \Longrightarrow 4\pi G\mu + \nabla^2\left(\int_0^p\frac{\d p'}{\mu(p')}\right) -2 \Omega^2= 0, \label{Laplace_Euler}
 \end{align}
 where $V_0$ is the constant of integration and $p$ the pressure.
 Applied to a function $f=f(r,\chi)$, the Laplace operator in the coordinates $(r,\chi,\varphi)$ reads
 \begin{align*}
   \nabla^2f &= \frac{\partial^2 f}{\partial r^2} + \frac{1}{r}\frac{\partial f}{\partial r}
          +\frac{1}{r^2}\frac{\partial^2 f}{\partial \chi^2}  \nonumber \\
  & \quad -\left(\frac{a}{\sigma} - r\cos\chi\right)^{-1}
          \left(\cos\chi\frac{\partial f}{\partial r} - \frac{\sin\chi}{r}\frac{\partial f}{\partial \chi}\right).
 \end{align*}

 Expanding \eqref{Laplace_Euler} in terms of $\sigma$ and requiring that
 the equations be satisfied for each power in $\sigma$ and each term in the Fourier
 expansion then results in ODEs for $\mu_{qk}(y)$ once an
 equation of state has been specified. These functions must
 be regular at the origin and chosen such that $\mu_{00}(0)=1$ and $\mu_{ik}(0)=0$ for all other
 $i$ and $k$ so as to be consistent with the choice $\mu(0,\chi)=\mu_\text c$. For $k=0$, this
 condition suffices to determine the function uniquely. For $k=1,2,\ldots,q$, the remaining constants
 in the solution of the ODEs are found by requiring that the pressure vanish at the surface. Demanding this
 for each of the coefficients in a Fourier expansion, provides $q+1$ equations for the remaining $q$ constants.
 The additional equation can be used to determine $\beta_{q0}$. It may come as a surprise that a term with $k=0$
 was included in the expansion here for $\rs$, since no such term is needed in the homogeneous case. The scale
 invariance mentioned toward the end of \cite{HP08} means that $a$ need never be specified in that case%
 \footnote{Polytropes with $n=1$, which will be treated shortly, also contain an interesting invariance.
  If $U(\mathbf x)$, $\mu(\mathbf x)=K\sqrt{p(\mathbf x)}$, $\Omega$ and $V_0$ are solutions to the Poisson
  and integrated Euler equations, then so are $\alpha U(\mathbf x)$, $\alpha \mu(\mathbf x)=K \sqrt{\alpha^2 p(\mathbf x)}$,
  $\sqrt{\alpha}\Omega$ and $\alpha V_0$, where $\alpha$ is an arbitrary scaling factor. This invariance
  is reflected in the fact that the value for $a$ is independent of $\mu_\text c$ for polytropes with $n=1$.}.
 Here, however, the first zero of $\mu_{00}$ determines the value of $a$, which can then be `corrected' order
 for order in $\sigma$ via the coefficients $\beta_{i0}$.

 Now that the density has been determined and the number of unknowns coincides with that of the
 homogeneous case, we can proceed as we did there, a detailed description of which can be found in \citet{HP08}.
 The freedom one has to choose the origin of the coordinate system, $r=0$, is used in requiring that it coincide
 with the centre of mass
 \begin{equation}\label{com}
  \int_0^{2\pi}\int_0^{\rs(\chi)} r^2\mu(r,\chi) \cos\chi\, \d r\,\d\chi =0.
 \end{equation}
 The potential in the vacuum region is determined from the Poisson integral via a rather involved
 procedure and reads
 \begin{equation}\label{U_out}
  U_\text{out} = -2\pi G\mu_\text c a^2\left(\sum_{l=1}^q a^{2l-1}\sigma^{-l} A_l I_l + o(\sigma^q)\right),
 \end{equation}
 where
 \begin{equation}
  I_l := \left(-\frac{1}{b}\frac{\d}{\d b} \right)^{l-1}\int_0^\pi
         \frac{\d\varphi}{\sqrt{b^2+\varrho^2+z^2-2b\varrho\cos\varphi}}.
 \end{equation}
 and the terms $A_l$ result from an expansion of the Poisson integral.

 Making use of expansion formul\ae\ for $I_l$, $U_\text{out}$ can be evaluated
 along the surface of the ring $\rs$ and expanded in terms of $\sigma$. The Euler
 equation then tells us that
 \begin{equation}\label{Euler}
  U_\text{out}(\rs) - \frac{1}{2}\Omega^2(a/\sigma-\rs\cos\chi)^2=V_0
 \end{equation}
 holds. Considering the $\sigma^q$ term and equating the coefficients of the $\cos(k\chi)$,
 $k=1,2,\ldots,q$ to zero then provides equations for determining the remaining unknowns.

 It should be noted that the expansion coefficients depend on powers of $\ln\sigma$ in general.
 Because $\lim_{\sigma\to 0}\sigma(\ln\sigma)^\alpha=0$ for all $\alpha$, this dependence does
 not pose a problem for the iteration scheme.

 \section{General Results to First Order}

 The approximation scheme described above allows us to draw certain conclusions even without
 specifying the equation of state, thus generalizing results that were published for polytropes
 by \citet{Ostriker64b}.  To leading order in $\sigma$, where nothing depends on the
 angle $\chi$, the ring (here a torus) is equivalent to an infinitely long cylinder, a problem
 that was studied by \cite{CF53, Ostriker64}. Upon introducing the pressure function
 \begin{align}
  h:= \int_0^p\frac{\d p'}{\mu(p')}
 \end{align}
 and expanding it as with $\mu$ in \eqref{mu_expansion}
 \begin{align}
  h(r,\chi)   &= G\mu_\text c a^2\left(\sum_{i=0}^q\sum_{k=0}^i h_{ik}(y)\cos(k\chi) \sigma^i+ o(\sigma^q)\right),
 \end{align}
 equation \eqref{Laplace_Euler} reads
 \begin{align}\label{h00}
  \left(\frac{\d^2}{\d y^2}+ \frac{1}{y}\frac{\d}{\d y}\right)h_{00} + 4\pi\mu_{00} = 0
 \end{align}
 to the lowest order in $\sigma$, since the integrated Euler equation \eqref{int_Euler} tells us that
 \begin{align}\label{Om_order}
  \Omega^2/G\mu_\text c = o(\sigma^2)
 \end{align}
 must hold, i.e.
 \begin{align}
  \Omega_0=\Omega_1=0.
 \end{align}

\begin{figure}
  \centerline{\includegraphics{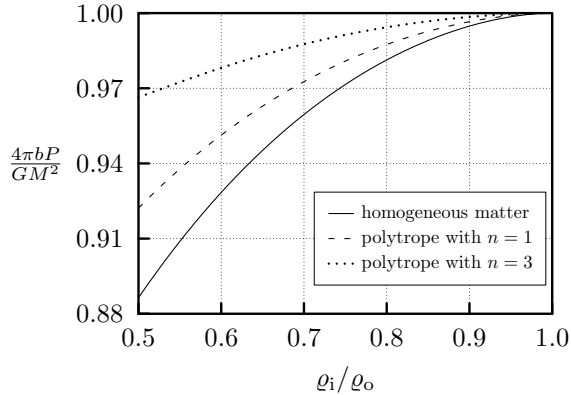}}
  \caption{Numerical examples demonstrating how $4\pi b P/GM^2$ tends to 1 in the thin ring
           limit for various equations of state, cf.\ equation \eqref{eq:P_M2}. \label{P_M2}}
 \end{figure}

 At the surface of the ring $r=\rs$, the pressure vanishes, corresponding
 to $h(r=\rs)=0$, and we thus find
 \begin{align}
  h_{00}(1)=0
 \end{align}
 and
 \begin{align}\label{eq:beta11}
  \beta_{11} = \left.-h_{11}\left(\frac{\d h_{00}}{\d y}\right)^{-1}\right|_{y=1}.
 \end{align}
 By multiplying \eqref{h00} by $\pi \mu_\text c a^2b y$ and integrating from 0 to 1, one finds that the
 mass $M$ to leading order can be related to the derivative of $h_{00}$ at the point $y=1$ according to
 \begin{align}\label{mass_h00}
  M = 4\pi^2 \mu_\text c a^2 b\int_0^1 \mu_{00}y\,\d y = -\pi a^3\mu_\text c/\sigma \left.\frac{\d h_{00}}{\d y}\right|_{y=1}.
 \end{align}
 A particularly interesting relation involving the square of the mass can be derived by considering
 the integral over the pressure $p$, which we first expand
 \begin{align}
  p(r,\chi)   &= G\mu_\text c^2 a^2\left(\sum_{i=0}^q\sum_{k=0}^i p_{ik}(y)\cos(k\chi) \sigma^i+ o(\sigma^q)\right).
 \end{align}
 To leading order, upon taking \eqref{h00} into account, the integral over $p$ reads
 \begin{align}
   P &:= 2\pi \int_0^{2\pi}\int_0^{r_\text s(\chi)} pr (b-r\cos\chi)\,\d r \,\d \chi,\intertext{which to leading order is}
\begin{split}\label{eq:P_M2}
   P & = 4\pi^2G\mu_\text c^2 a^4 b\int_0^1 p_{00} y\,\d y \\
     & = -2\pi^2G\mu_\text c^2 a^4 b\int_0^1 \frac{\d p_{00}}{\d y} y^2\,\d y \\
     &= -2\pi^2G\mu_\text c^2 a^4 b \int_0^1 \mu_{00} \frac{\d h_{00}}{\d y} y^2\,\d y\\
     &= 8\pi^3 G \mu_\text c^2 a^4 b \int_0^1 \mu_{00}y \left(\int_0^y \mu_{00}  y'\,\d y'\right)  \d y\\
     &= 4\pi^3 G \mu_\text c^2 a^4 b \left(\int_0^1 \mu_{00}y\, \d y\right)^{\!2}\\
     &= \frac{GM^2}{4\pi b}.
  \end{split}
 \end{align}
 Numerical examples demonstrating how $4\pi b P/GM^2$ approaches 1 in the thin ring limit
 for various equations of state can be found in Fig.~\ref{P_M2}.

\begin{figure}
  \centerline{\includegraphics{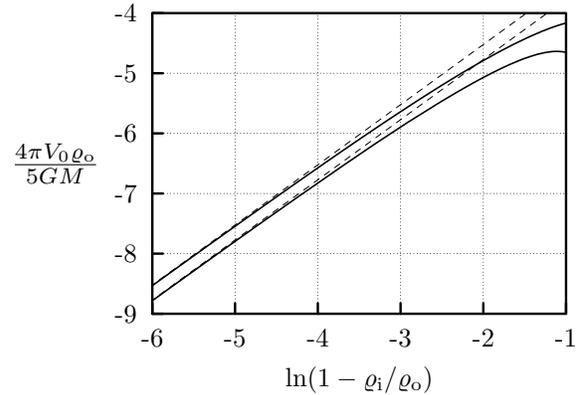}}
  \caption{Numerical ring sequences (solid lines) for homogeneous matter (upper curve)
           and polytropes with $n=5$ (lower curve) are plotted for rings approaching the thin ring limit.
           The asymptotic behaviour as given by equation \eqref{b_V_0_M} is indicated by the dashed lines.\label{n1n5}}
 \end{figure}

 The terms from the expansion \eqref{U_out} of the potential in the vacuum that play a role
 up to first order are
 \begin{align}
  A_1 &= 2\int_0^1 \mu_{00} y\,\d y = \frac{M\sigma}{2\pi^2 \mu_\text c a^3}\intertext{and}
  A_2 &= \frac{gM\sigma^2}{2\pi^2\mu_\text c a^3},\qquad g:=-\frac{\pi^2\mu_\text c a^3}{M\sigma}\int_0^1 \mu_{00} y^3\, \d y.
  \label{def_g}
 \end{align}
 The coefficient $\Omega_2$ from the expansion of the square of the angular velocity is
 \begin{align}\label{eq:Om2}
  \Omega_2 &= A_1(1+\lambda-2\beta_{11}) + 2A_2/\sigma\intertext{with} \lambda&:=\ln(8/\sigma) -2,
 \end{align}
 and the constant of integration from the Euler equation is
 \begin{align}
  \label{V0}
  \begin{split}
   V_0 &= \left.\left(U_\text{out} - \frac{1}{2}\Omega^2\varrho^2\right)\right|_{r=\rs}\\
      &= -\frac{GM}{2\pi b}\left(\frac{5\lambda+9}{2} +g -\beta_{11} \right).
  \end{split}
 \end{align}
 The term $g-\beta_{11}$ appearing in the above equation can be treated further by considering
 the rotational energy $T$ and potential energy $W$ and making use of the virial identity
 \begin{align}\label{VI_n1}
   \begin{split}
    0 &= 3P+ 2T+W \\
      &= 3P+ \frac{GM^2}{\pi b}\left(g-\beta_{11}-\frac{1}{2} \right)\\
      & \qquad   - 2\pi^2 G\mu_\text c^2 a^4 b\int_0^1 \mu_{00} h_{00}y\, \d y.
  \end{split}
 \end{align}
 By restricting ourselves to the polytropic equation of state (see \eqref{poly}), we
 can rewrite the above integral to read
 \begin{align}
  \begin{split}
    \int_0^1 \mu_{00}h_{00}y\, \d y &= \frac{(n+1)K\mu_\text c^{1/n-1}}{Ga^2}\int_0^1 \mu_{00}^{1+1/n} y\, \d y\\
    &=(n+1) \int_0^1 p_{00} y\, \d y
      =\frac{(n+1)M^2}{16\pi^3 \mu_\text c^2 a^4 b^2},
  \end{split}
 \end{align}
 where the last step follows from \eqref{eq:P_M2}. Putting this expression into \eqref{VI_n1}
 and using \eqref{eq:P_M2} again then yields
 \begin{align}\label{g_minus_beta11}
  g-\beta_{11} = \frac{n-1}{8}.
 \end{align}
 Taking into account $1-\varrho_\text i/\varrho_\text o = 2\sigma$, which holds to leading order,
 we can use \eqref{eq:Om2} to write
 \begin{align}\label{b_Om2_M}
  \frac{2\pi b^3 \Omega^2}{GM}+\ln\left(1-\frac{\varrho_\text i}{\varrho_\text o}\right) \to \frac{n-5}{4} + \ln 16
 \end{align}
 and \eqref{V0} can be written as
 \begin{align}\label{b_V_0_M}
  \frac{4\pi b V_0}{5GM}-\ln\left(1-\frac{\varrho_\text i}{\varrho_\text o}\right) \to \frac{5-n}{20} - \ln 16
 \end{align}
 for polytropes in the thin ring limit. Similar equations can be derived for $J$ (angular momentum), $T$, $P$ and via the
 virial identity for $W$ \citep[see][]{Ostriker64b}. These equations also hold for homogeneous bodies ($n=0$), as was
 shown in \cite{HP08}. Numerical examples demonstrating the behaviour \eqref{b_V_0_M} are provided in Fig.~\ref{n1n5}.

 \section{Mass Density for Polytropes at the Zeroth Order}

 The polytropic equation of state is
 \begin{equation}\label{poly}
  p = K\mu^{1+1/n}.
 \end{equation}
 For large/small polytropic indices $n$, the equation is referred to as `soft'/`stiff' and as $n$ tends to zero,
 $\mu$ tends to a constant. From now on, we shall use the terms `homogeneous matter' and `$n=0$' interchangeably.
 For polytropes, \eqref{Laplace_Euler} becomes
 \begin{align}\label{Laplace_Euler_polytrope}
  4\pi G\mu + K(n+1)\nabla^2\left(\mu^{1/n}\right) -2 \Omega^2= 0.
 \end{align}
 Instead of our coordinate $y$, we are now going to make use of a new dimensionless radial coordinate,
 applicable to polytropes
 \begin{align}
  x := \frac{G^\frac{1}{2}\mu_\text c^\frac{n-1}{2n}}{K^\frac{1}{2}}  r.
 \end{align}
 To lowest order in $\sigma$, and upon introducing
 \begin{align}
  \tilde\mu:=\mu^{1/n}
 \end{align}
 and the expansion
 \begin{align}
  \tilde\mu   &= \mu_\text c^{1/n}\left(\sum_{i=0}^q\sum_{k=0}^i \tilde\mu_{ik}(x)\cos(k\chi) \sigma^i + o(\sigma^q)\right),
 \end{align}
 equation \eqref{Laplace_Euler_polytrope} reads, cf.\ \eqref{h00},
 \begin{equation}\label{Lane-Emden}
  \left(\frac{\d^2}{\d x^2}+ \frac{1}{x}\frac{\d}{\d x}\right)\tilde\mu_{00} 
 + \frac{4\pi}{n+1}\tilde\mu_{00}^{ n}= 0.
 \end{equation}
 This equation is sometimes referred to as one of the {\sl generalized Lane-Emden equations (of the first kind)} and
 solutions to it have been derived and studied in e.g.\ \citet{GH00}. No solutions other than for $n=1$ have been
 found for our particular parameters in closed-form and a discussion using symmetry transformations suggests that they do not
 exist, \citep{Goenner01}. We thus concentrate in the next section on the special case $n=1$.

 \section{Analytic Solution for Polytropes with $\bmath{\lowercase{n}=1}$}
  \subsection{The Zeroth Order: $\bmath{\sigma^0}$}
  We rewrite \eqref{Lane-Emden} for $n=1$, remembering that now $\tilde\mu=\mu$,
  \begin{equation}
   \left(\frac{\d^2}{\d x^2}+ \frac{1}{x}\frac{\d}{\d x}\right)\mu_{00} + 2\pi\mu_{00}= 0
  \end{equation}
  and can immediately write down the general solution
  \begin{equation}
   \mu_{00} = C_1 J_0(\sqrt{2\pi} x) + C_2 Y_0(\sqrt{2\pi} x),
  \end{equation}
  where $J_n$ is a Bessel function (of the first kind) and $Y_n$ a Neumann function (also called a
  Bessel function of the second kind), see e.g.\ \citet*{PBM90b}. The condition
  $\mu(r=0,\chi)=\mu_\text c$ tells us that $C_1=1$
  and $C_2=0$. The first positive zero of $J_0$ determines value
  for $a$ from \eqref{rs}. We refer to the $k$th positive zero of the $n$th Bessel function as $j_{nk}$
  and can then write
  \begin{equation}
   a x/r=: \bar{a}= j_{01}/\sqrt{2\pi} = 0.959\ldots
  \end{equation}

 \subsection{The First Order: $\bmath{\sigma^1}$}
  The unknown quantities we have to solve for are $\mu_{10}(x)$, $\mu_{11}(x)$, $\beta_{10}$, $\beta_{11}$, and $\Omega_2$.
  From \eqref{Laplace_Euler}, one finds the differential equations
  \begin{align}
   &\left(\frac{\d^2}{\d x^2}+ \frac{1}{x}\frac{\d}{\d x}\right)\mu_{10} + 2\pi\mu_{10}= 0 \intertext{and}
   &\left(\frac{\d^2}{\d x^2}+ \frac{1}{x}\frac{\d}{\d x}\right)\mu_{11} + \left(2\pi-\frac{1}{x^2}\right)\mu_{11}=
      \frac{1}{ \bar a}\frac{\d\mu_{00}}{\d x}.
  \end{align}
  Considering only solutions that vanish at the point $x=0$, so as to maintain our choice $\mu(0)=\mu_\text c$,
  we find
  \begin{align}
   \mu_{10} &= 0 \intertext{and}
   \mu_{11} &= C_3 J_1 + \frac{1}{2 j_{01}}\left(\sqrt{2\pi}x J_0 - J_1\right),
  \end{align}
  where the argument of the Bessel function is always $\sqrt{2\pi}x$ unless otherwise specified.
  The requirement that the density vanish at the surface of the rings determines
  \begin{align}
   \beta_{10}=0
  \end{align}
  and relates the constant $C_3$ to the surface function
  \begin{equation}
   C_3 = \frac{1+2 j_{01}^2 \beta_{11}}{2j_{01}}.
  \end{equation}
  The constant $\beta_{11}$ is determined by stipulating that the centre of mass coincide with the
  point $(\varrho=b,z=0)$ as in \eqref{com}
  \begin{equation}
   \beta_{11} = \frac{4-j_{01}^2}{4j_{01}^2}.
  \end{equation}
  Recalling the definition $\lambda := \ln\frac{8}{\sigma}-2$ one finally obtains
  \begin{equation}
   \Omega_2 = \frac{2J_1(j_{01})(\lambda+1)}{j_{01}}
  \end{equation}
  from \eqref{Euler}.

 \subsection{The Second Order: $\bmath{\sigma^2}$}
  To second order, the unknown quantities that have to be solved for are $\mu_{20}(x)$, $\mu_{21}(x)$, $\mu_{22}(x)$,
  $\beta_{20}$, $\beta_{21}$, $\beta_{22}$, and $\Omega_3$.

  The ODEs describing the mass density now read
  \begin{align}
  \begin{split}
   &\left(\frac{\d^2}{\d x^2}+ \frac{1}{x}\frac{\d}{\d x}\right)\mu_{20} + 2\pi\mu_{20}= \pi\Omega_2
      \\& \qquad+ \frac{1}{2 \bar a}\left(\frac{\d\mu_{11}}{\d x} - \frac{\mu_{11}}{x}\right)
               + \frac{x}{2 \bar a^2}\frac{\d\mu_{00}}{\d x},
  \end{split}\\
   &\left(\frac{\d^2}{\d x^2}+ \frac{1}{x}\frac{\d}{\d x}\right)\mu_{21} + \left(2\pi-\frac{1}{x^2}\right)\mu_{21}=
      \frac{1}{ \bar a}\frac{\d\mu_{10}}{\d x}=0\intertext{and}
   \begin{split}
   &\left(\frac{\d^2}{\d x^2}+ \frac{1}{x}\frac{\d}{\d x}\right)\mu_{22} + \left(2\pi-\frac{4}{x^2}\right)\mu_{22}=
      \\& \qquad+ \frac{1}{2 \bar a}\left(\frac{\d\mu_{11}}{\d x} - \frac{\mu_{11}}{x}\right) 
               + \frac{x}{2 \bar a^2}\frac{\d\mu_{00}}{\d x}.
   \end{split}
  \end{align}
  The solutions vanishing at $x=0$ are
  \begin{align}
   \mu_{20} &= \frac{\Omega_2}{2}\left(1-J_0\right) +
      \frac{1}{8}\left[\frac{3\pi x^2}{j_{01}^2}J_0 + \sqrt{2\pi}x \left(\frac{1}{j_{01}^2}-\frac{1}{2}\right)J_1 \right], \\
   \mu_{21} &= C_4 J_1 \intertext{and}
  \begin{split}
   \mu_{22} &= C_5 J_2 
    + \frac{1}{4}\left[\left(\frac{5}{j_{01}^2} + \frac{3\pi x^2}{2 j_{01}^2} -\frac{1}{2}\right)J_2 \right.\\
    & \qquad +\left. \frac{\sqrt{\pi}x}{\sqrt{2}}\left(\frac{5}{j_{01}^2}-\frac{1}{2}\right)J_1\right].
   \end{split}
  \end{align}

 The constants $C_4$ and $C_5$ can be related to the surface function by requiring that $\mu(\rs)=0$ hold independently
 for the coefficients in front of $\cos\chi$ and $\cos 2\chi$. The result is
 \begin{align}
  C_4 &= j_{01} \beta_{21} \intertext{and}
  C_5 &= \frac{1}{2}\left(-j_{01}^2 \beta_{22} -\frac{j_{01}^2}{64}+\frac{3}{8} -\frac{11}{4j_{01}^2}\right).
 \end{align}
 Requiring the same of the coefficient in front of $\cos 0\chi$ gives
 \begin{equation}
  \beta_{20} = -\frac{4}{j_{01}^4} + \frac{\lambda+1}{j_{01}^2} -\frac{1}{64}.
 \end{equation}
 Evaluating \eqref{com} tells us that
 \begin{equation}
  \beta_{21} = 0 \Longrightarrow \mu_{21}=0.
 \end{equation}
 The values for the remaining constants follow from \eqref{Euler}:
 \begin{align}
  \Omega_3= 0
 \end{align}
 and
 \begin{align}
  \beta_{22}  = \frac{1}{4j_{01}^4} + \frac{5(\lambda+3)}{2j_{01}^2} + \frac{1}{64}.
 \end{align}

\subsection{The Third Order: $\bmath{\sigma^3}$}

 The third order is the final one to be presented here, but the iterative scheme can
 be applied up to arbitrary order assuming that one is able to solve the differential
 equations for the mass density and perform the necessary integrals. The ODEs that result for this order are
 \begin{align}
   &\left(\frac{\d^2}{\d x^2}+ \frac{1}{x}\frac{\d}{\d x}\right)\mu_{30} + 2\pi\mu_{30}= 0,\\
   \begin{split}
   &\left(\frac{\d^2}{\d x^2}+ \frac{1}{x}\frac{\d}{\d x}\right)\mu_{31} + \left(2\pi-\frac{1}{x^2}\right)\mu_{31}=
        \frac{3x^2}{4 \bar a^3}\frac{\d\mu_{00}}{\d x}
      \\& \quad  + \frac{3}{4 \bar a^2}\left(x\frac{\d}{\d x} + 1 \right)\mu_{11}
      + \frac{1}{ \bar a}\left[\frac{\d\mu_{20}}{\d x} + \left(\frac{1}{2}\frac{\d}{\d x}+ \frac{1}{x}\right)\mu_{22}\right],
   \end{split}\\
   &\left(\frac{\d^2}{\d x^2}+ \frac{1}{x}\frac{\d}{\d x}\right)\mu_{32} + 2\left(\pi -\frac{2}{x^2}\right)\mu_{32}= 0
   \intertext{and}
   \begin{split}
   &\left(\frac{\d^2}{\d x^2}+ \frac{1}{x}\frac{\d}{\d x}\right)\mu_{33} + \left(2\pi-\frac{9}{x^2}\right)\mu_{33}=
       \frac{x^2}{4 \bar a^3}\frac{\d\mu_{00}}{\d x}
      \\& \quad + \frac{1}{4 \bar a^2}\left(x\frac{\d}{\d x} - 1 \right)\mu_{11}
       + \frac{1}{ \bar a}\left(\frac{1}{2}\frac{\d}{\d x}- \frac{1}{x}\right)\mu_{22}.
   \end{split}
 \end{align}
 The solutions to these equations obeying the requirement $\mu_{qk}(0)=0$ are
 \begin{align}
  \mu_{30}&=0,\\
  \begin{split}
   \mu_{31}&=
    \left(C_6
       -\frac {\left( \lambda+1 \right)J_1\!\left(j_{01} \right) }{2j_{01}^{2}}
       -{\frac {9}{64}} {\frac {\pi  \left( j_{01}^{2}-8\right) {x}^{2}}{j_{01}^{3}}}\right.\\
     &-\left.\frac {j_{01}^{4} + 16j_{01}^{2}(5 \lambda +4) -16}{256 j_{01}^{3}} \right) J_1\\
     & + \left( \frac {\sqrt {\pi }\left( \lambda+1 \right) J_1\!\left(j_{01} \right)x }{\sqrt{2}j_{01}^{2}}
        -{\frac {15}{32}} {\frac {\sqrt {2}{\pi }^{3/2}{x}^{3}}{j_{01}^{3}}} \right.\\
     &+\left.\frac {\sqrt {2\pi }\left(j_{01}^{4} + 16j_{01}^{2}(5\lambda+4)-16
         \right) x}{256 j_{01}^{3}} \right) J_2,
  \end{split}\\
  \mu_{32}&=C_7 J_2\intertext{and}
  \begin{split}
    \mu_{33}&=
    \frac{\pi  {x}^{2}\left(16j_{01}^{2}(5\lambda+2)-40 \pi {x}^{2}+272+j_{01}^{4}\right)}{512j_{01}^{3}} J_1 \\
    &\left(-\frac {5{\pi }^{2}{x}^{4}}{64 j_{01}^{3}}
      +\frac {\pi \left(j_{01}^{4}+8j_{01}^{2}(10\lambda+7)+16\right)x^2}{512 j_{01}^{3}}\right.\\
    &-\left.\frac {3\left(j_{01}^{4}+16j_{01}^{2}(5\lambda + 3)+208\right)}
       {256 j_{01}^{3}} + C_8\right)J_3.
  \end{split}
 \end{align}

 The constants $\beta_{30}$, $C_6$, $C_7$ and $C_8$ are determined by requiring that
 $\mu(r=r_\text s)=0$ hold independently for the coefficients in front
 of the $\cos 0\chi$, $\cos \chi$, $\cos 2\chi$ and $\cos 3\chi$ terms. The result is
 \begin{align}
  \begin{split}
   \beta_{30}&= 0,\\
   C_6&=
   \left( \frac{\lambda+1}{4} -\frac{3\left( \lambda+1 \right) }{2j_{01}^{2}} \right) J_1\!\left(j_{01} \right)
     -{\frac {j_{01}^{3}}{512}}\\
   &\qquad + j_{01}   \left( \frac{5}{32}\lambda+\beta_{31}+\frac{39}{256} \right)\\
   &\qquad -\frac{ 14 \lambda + 9 }{32 j_{01}}
    +\frac { 40 \lambda + 37}{16 j_{01}^{3}},
  \end{split}\\
  C_7&=0\intertext{and}
  \begin{split}
   C_8&= \frac{1}{j_{01}^2-8}\Bigg[-{\frac {j_{01}^{5}}{1536}}
     - \left({\frac {5}{32}} \lambda +\beta_{33}+{\frac {77}{768}}\right) j_{01}^{3}\\
   & \quad + \left( {\frac {35}{16}} \lambda+{\frac {11}{8}} \right) j_{01}
     - \frac{1}{ {j_{01}}}\left( 10 \lambda+{\frac {181}{48}} \right)
   -{\frac {119}{6 j_{01}^{3}}}\Bigg].
  \end{split}
 \end{align}
 Equation \eqref{com} yields
 \begin{align}
   \beta_{31}= {\frac {9}{32}} \lambda + {\frac {31}{128}}
     - \frac{1}{ j_{01}^{2}}\left({\frac {13}{8}} \lambda + {\frac {5}{16}} \right)
     -\frac{1}{j_{01}^{4}}\left(\frac{5}{2} \lambda + 9\right)
 \end{align}
 and \eqref{Euler} gives
 \begin{align}
  \beta_{32}&=0,\\
  \begin{split}
   \Omega_4&=-2 {\frac { \left(\lambda +1\right) ^{2} }{j_{01}^{2}}} J_1\!\left(j_{01} \right)^2
    + \Bigg[ \frac{1}{32}\left(\lambda +1 \right) j_{01}\\
   & \quad +{\frac{1}{j_{01}}}\left({\lambda}^{2}+\frac{3}{8} \lambda-\frac{1}{2}\right)\\
   & \quad + \frac{1}{j_{01}^3}\left( \frac{11}{2}\lambda + {\frac {17}{4}}\right)  \Bigg] J_1\!\left(j_{01}\right)
  \end{split}\intertext{and}
  \beta_{33}&={\frac {5}{64}}- \frac{1}{8 j_{01}^{2}}\left(5\lambda +9 \right)
    +\frac{1}{j_{01}^{4}} \left(\frac{5}{2}\lambda+2\right).
 \end{align}

 \subsection{Physical Parameters}

\begin{figure}
  \centerline{\includegraphics{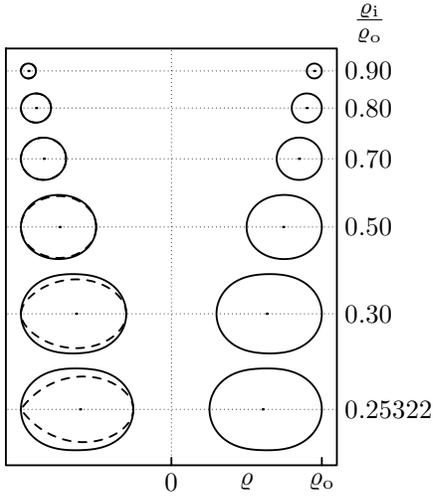}}
  \caption{Meridional cross-sections of polytropic rings ($n=1$) with varying radius ratio are shown to third order
           (solid lines) in comparison to numerical results (dashed lines) for the same radius
           ratio. At the value $\varrho_\text i/\varrho_\text o=0.25322\ldots$, the rings reach a mass-shedding
           limit, as is evident from the numerical cross-section.\label{fig_rs}}
 \end{figure}

 The shape of the rings to third order that results from equation \eqref{rs}
 is compared to numerical results of the corresponding radius ratio in Fig.~\ref{fig_rs}.
 For thin rings, the numerical and third order curves are indistinguishable. As the radius
 ratio is decreased, the numerical results show that the outer edge becomes pointier, right up
 to the mass-shedding limit for the value $\varrho_\text i/\varrho_\text o=0.25322\ldots$.
 For such a ring, a fluid particle rotating at the outer rim in the equatorial plane has a
 rotational frequency equal to the Kepler frequency, meaning that it is kept in balance by
 the gravitational and centrifugal forces alone -- the force arising from the pressure gradient
 vanishes. The shape of the ring with the cusp that forms for mass-shedding configurations
 is not well represented by a small number of terms in our Fourier series.

 Using the results of the last subsection, we write down expressions for various
 physical parameters and can use them to verify that the virial
 identity $3P+2T+W=0$ is satisfied to each order in $\sigma$.
 For convenience, we first introduce dimensionless quantities, valid
 for any polytropic index $n>0$:
 \begin{align}
  \begin{split}
   \frac{\bar{M}}{M}= \frac{G^\frac{3}{2}\mu_\text c^{\frac{n-3}{2n}}}{K^\frac{3}{2}},
       \quad
   \frac{\bar{J}}{J}= \frac{G^2\mu_\text c^\frac{2n-5}{2n}}{K^\frac{5}{2}},
      \quad \frac{\bar{\Omega}}{\Omega}= \frac{1}{G^\frac{1}{2}\mu_\text c^\frac{1}{2}},\\
   \frac{\bar{P}}{P}= \frac{\bar{T}}{T}= \frac{\bar{W}}{W}= \frac{G^\frac{3}{2}\mu_\text c^\frac{n-5}{2n}}{K^\frac{5}{2}},
       \quad
   \frac{\bar{b}}{b} = \frac{\bar{\varrho}}{\varrho}=\frac{G^\frac{1}{2}\mu_\text c^\frac{n-1}{2n}}{K^\frac{1}{2}},
  \end{split}
 \end{align}
 where $J$ refers to the angular momentum. Up to and including third order one finds
 \begin{align}
   \begin{split}
   &\bar M= \,
     {\frac {\sqrt {2\pi }j_{01}^{2}J_1\!\left( j_{01} \right) }{\sigma}}\\
   &\qquad +\bigg[\frac{\sqrt {2\pi }}{64} \left(j_{01}^{4} +  28 j_{01}^{2}+32 j_{01}^{2}\lambda-16 \right)
            J_1\!\left( j_{01} \right)\\
   &\qquad  -\sqrt {2\pi } j_{01}  \left( \lambda+1 \right)
             J_1\!\left(j_{01} \right)^2  \bigg]  \sigma,
  \end{split}\\
  \begin{split}
   &\bar J= \,
       \sqrt {{\frac {J_1\!\left(j_{01} \right)\left( \lambda+1 \right) }{ j_{01}}}}J_1\!\left(j_{01} \right)j_{01}^{2}
     \Bigg\{\frac{j_{01}^{2}}{ {\sigma}^2}\\
     &\qquad     -\frac{3}{2}\left( \lambda+1 \right)j_{01} J_1\!\left(j_{01} \right)\\
     &\qquad + {\frac {1}{128\left( \lambda+1 \right)}}\big[3 j_{01}^{4}(\lambda+1)\\
     &\qquad +j_{01}^{2}(96 {\lambda}^{2}+324 \lambda+232) -624\lambda-664 \big]\Bigg\},
  \end{split}\\
  \begin{split}
   &\bar P=\,
      \frac{\sqrt {2\pi}}{2}j_{01} J_1\!\left(j_{01} \right)^2\bigg\{
      \frac{j_{01}^{2}}{\sigma}\\
      &\qquad+\frac {1}{320}
       \Big[ -640 j_{01} \left( \lambda+1 \right)J_1\!\left(j_{01} \right)\\
      &\qquad+10\left( j_{01}^{4}-8 j_{01}^{2}+128\lambda+136 \right) \Big] \sigma \bigg\},
  \end{split}\\
  \begin{split}
   &\bar T=\,
      \frac{\sqrt {2\pi}}{2} J_1\!\left(j_{01} \right)  ^2 { j_{01}}\Bigg\{
      \frac {1}{\sigma} \left( \lambda+1 \right) j_{01}^{2}\\
    &\qquad\Big[ - 2\left( \lambda+1 \right) ^{2}J_1\!\left( j_{01} \right)
       { j_{01}}+{\frac {1}{32}}\Big(j_{01}^{4}(\lambda+1)\\
    &\qquad + 2j_{01}^{2}(8\lambda+9)(2\lambda+3)-112\lambda-132\Big)\Big] \sigma \Bigg\}
  \end{split}\intertext{and}
  \begin{split}
   &\bar W=\,
     \sqrt{2\pi}j_{01}J_1\!\left(j_{01} \right)^2\Bigg\{
     -{\frac{j_{01}^{2}}{2\sigma} \left( 2 \lambda+5 \right)}\\
    &\qquad \Bigg[ { j_{01}} \left( 2 \lambda+5 \right)  \left( \lambda+1 \right)J_1\!\left(j_{01} \right)\\
    &\qquad -{\frac {1}{64}}\Big(j_{01}^{4}(2 \lambda+5)+ 4 j_{01}^{2}(16{\lambda}^{2}+42 \lambda+21)\\
    &\qquad\qquad+160\lambda +144\Big)\Bigg] \sigma\Bigg\}.
  \end{split}
 \end{align}
 In the derivation of the above expressions for $\bar P$ and $\bar W$, we have made use
 of the identity
 \begin{align}\label{F_identity}
   20\, _2\!F_3\!\left(\frac{3}{2},\frac{3}{2};2,2,\frac{5}{2};-j_{01}^{2} \right)
   = 3j_{01}^{2}\, _2\!F_3\!\left(\frac{5}{2},\frac{5}{2};3,3,\frac{7}{2};-j_{01}^{2} \right)
 \end{align}
 for the Gauss hypergeometric function
 \begin{align}
  \begin{split}
 & _p\!F_q(a_1,a_2,\ldots,a_p;b_1,b_2,\ldots,b_q;z)\\
 &\qquad :=\sum_{k=0}^\infty
    \frac{{(a_1)}_k \cdot{(a_2)}_k \cdots {(a_p)}_k}{{(b_1)}_k \cdot{(b_2)}_k \cdots {(b_q)}_k}\frac{z^k}{k!}
  \end{split}\intertext{with the Pochhammer bracket}
 & {(a)}_k:= a(a+1)\cdots(a+k-1), \qquad {(a)}_0:=1.\nonumber
 \end{align}
 A proof of \eqref{F_identity} can be found in Appendix~\ref{app:F}.

\begin{figure}
  \centerline{\includegraphics{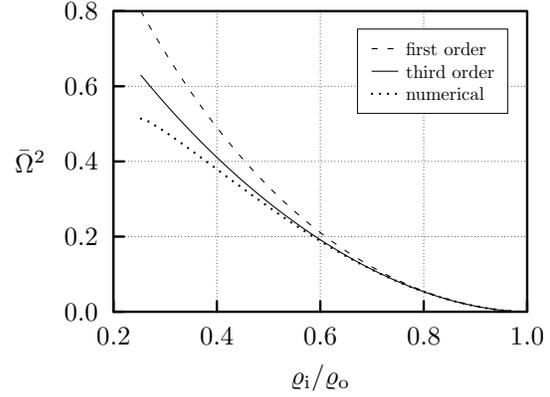}}
  \caption{The square of the dimensionless angular velocity is plotted versus
           the radius ratio for rings with polytropic index $n=1$.\label{Om2_n_poly_1}}
 \end{figure}

\begin{figure}
  \centerline{\includegraphics{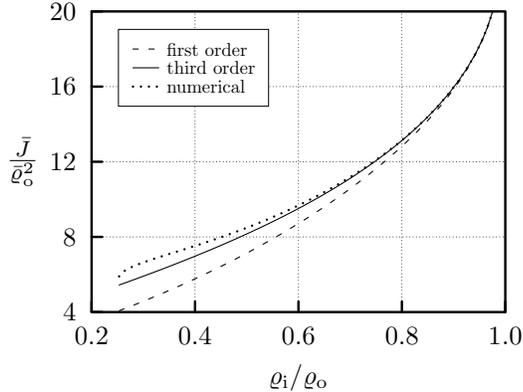}}
  \caption{The dimensionless angular momentum divided by the square of the outer radius is plotted versus
           the radius ratio for rings with polytropic index $n=1$. This quantity as a function of
           radius ratio tends logarithmically to infinity.\label{J_n_poly_1}}
 \end{figure}

 In order to gauge the accuracy of the expressions listed above, some of them are
 plotted to first and third order in comparison to numerical values in Figs~\ref{Om2_n_poly_1}--\ref{J_n_poly_1}.
 The accuracy of the numerical values is high enough so as to render the corresponding curve
 indistinguishable from the `correct' one and is plotted in its entirety, i.e.\ from the thin
 ring limit right up to the mass-shedding limit.
 The curves to first and third order were drawn by taking the expression for $\bar{\varrho}_\text i$,
 $\bar{\varrho}_\text o$, $\bar{\Omega}^2$, $\bar{M}$ and $\bar{J}$ to first and third order respectively,
 inserting a numerical value for $\sigma$ and then taking the appropriate combination of these numbers.
 One finds in all three plots that the third order brings a marked improvement as compared to the first
 one, but that the behaviour near the mass-shedding limit is not particularly well represented.
 \begin{figure}
  \centerline{\includegraphics{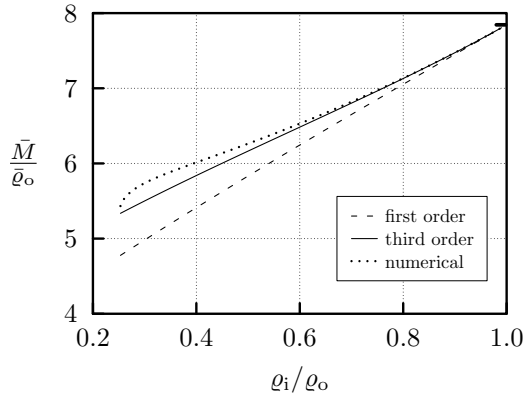}}
  \caption{The dimensionless mass divided by the outer radius is plotted versus
           the radius ratio for rings with polytropic index $n=1$. This quantity tends to
           the value $\bar{M}/\bar{\varrho}_\text o=2\pi j_{01} J_1(j_{01})=7.84\ldots$
           in the thin ring limit, which is marked by a tick.}
 \end{figure}

 \section{Solution for an Arbitrary Polytropic Index}

 As was mentioned above, our generalized Lane-Emden equation for $\tilde\mu_{00}$ can only be solved
 in closed-form for $n=1$. For other polytropic indices, the iterative method presented here was
 applied with the help of numerics. By describing the unknown density terms $\tilde\mu_{ik}$ by
 Chebyshev polynomials and expanding all the quantities involved in terms of $\lambda$, equations
 can be formulated for purely numerical coefficients. The equations of the approximation scheme described in
 Section~\ref{approx} must be fulfilled, whereby the ODEs for $\tilde\mu_{ik}$
 are evaluated at collocation points of the Chebyshev polynomials. In general, the density functions
 $\tilde\mu_{00}$ are not analytic at $x=\bar a$, meaning that high order polynomials may be necessary to find a
 good approximation of the function desired. We none the less chose this method, since the equations
 involve integrals over the density for which one end-point of integration contains the unknown surface
 function $\rs$, making their polynomial representation particularly useful.

 If one is only interested in determining $\bar a$, $\beta_{11}$ and $\Omega_2$, then it is not
 necessary to combine such numerical and algebraic techniques and one can choose
 any numerical method for solving the ODEs. One begins by solving equation \eqref{Lane-Emden}
 numerically for the desired polytropic index $n$, prescribing the `initial conditions'
 $\tilde\mu_{00}(0)=1$ and $\left.\frac{\d}{\d x}\tilde\mu_{00}\right|_{x=0}=0$.
 For spherical polytropic fluids, a surface of vanishing pressure is known to
 exist only for $n<5$, where the surface for $n=5$ extends out to infinity.
 The situation for polytropic rings is quite different -- it seems that arbitrary polytropic
 indices are possible! Numerical solutions to \eqref{Lane-Emden} indicate that the
 density function $\mu_{00}$ indeed falls to zero for large $n$.
 The value of $x$ at the first
 zero of the solution is $\bar a$. One then proceeds to solve equation
 \begin{align}
  \left(\frac{\d^2}{\d x^2} + \frac{1}{x}\frac{\d}{\d x} -\frac{1}{x^2}\right)\tilde\mu_{11}
   + \frac{4\pi n}{n+1}\tilde\mu_{00}^{n-1}\tilde\mu_{11}=\frac{1}{\bar a}\frac{\d\tilde\mu_{00}}{\d x}
 \end{align}
 for $\tilde\mu_{11}$ with the condition $\tilde\mu_{11}(0)=0$ and where $\left.\frac{\d}{\d x}\tilde\mu_{11}\right|_{x=0}$
 has to be chosen so as to fulfil the centre of mass condition $\eqref{com}$ to first order, which reads
 \begin{align}
  \begin{split}
    0 &= \int_0^{\bar a} \mu_{11} x^2\, \d x =  n\int_0^{\bar a} \tilde\mu_{00}^{n-1}\tilde\mu_{11} x^2\, \d x\\
      \Longrightarrow 0&= \bar a^2\left.\frac{\d\tilde\mu_{11}}{\d x}\right|_{x=\bar a} - \bar a\tilde\mu_{11}(\bar a)
       + \frac{2}{\bar a}\int_0^{\bar a} \tilde\mu_{00} x\, \d x.
  \end{split}
 \end{align}
 The constant $\beta_{11}$ can then be found using equation \eqref{eq:beta11}, which now reads
 \begin{align}
  \beta_{11} & = -\frac{\tilde\mu_{11}(\bar a)}{\bar a}\left.\left(\frac{\d\tilde\mu_{00}}{\d x} \right)^{-1}\right|_{x=\bar a},
 \end{align}
 and $\Omega_2$ is taken from \eqref{b_Om2_M}.
 The behaviour of these coefficients as they depend on the polytropic index $n$ can be found in Table~\ref{beta11_Om2}.
 The table suggests that $\bar a\to \infty$ and $\Omega_2\to 0$ exponentially in $n$ for $n\to\infty$, which is indeed
 known to hold \citep{Ostriker64,Ostriker64b}. The
 behaviour of the specific kinetic energy of a particle in the ring, proportional to $\bar a^2 \Omega_2$ to
 leading order, will be discussed in the next subsection together with the behaviour of $\beta_{11}$
 for large $n$.
 \begin{table}
  \centering
  \caption{The values of expansion coefficients for the surface function $\rs$ and for the squared angular
           velocity are provided up to first order for different polytropic indices $n$. The value of
           $\bar a$ for $n=0$ can be found by solving \eqref{Lane-Emden} with $n=0$ and the conditions
           $\frac{\d}{\d x}\tilde\mu_{00}|_{x=0}=0$ and $\tilde\mu_{00}(0)=1$ and then locating
           the first zero of $\tilde\mu_{00}$.\label{beta11_Om2}}
  \begin{tabular}{cccc}\toprule
    $n$ & $\bar{a}$                    & $\beta_{11}$ & $\Omega_2$\\ \midrule
     0  & $1/\sqrt{\pi}\approx 0.5642$ & 0            & $\lambda+3/4$ \\
    0.5 & $0.7566$                     & $-0.03537$    & $0.6371\lambda+0.5575$\\
     1  & $0.9594$                     & $-0.07708$    & $0.4318(\lambda+1)$\\
     2  & $1.427$                      & $-0.1731$     & $0.2169\lambda+0.2711$\\
     5  & $3.750$                      & $-0.5118$     & $0.03614\lambda+0.07228$\\
     10 & $15.18$                      & $-1.126$      & $(2.401\lambda+7.804)\times 10^{-3}$\\
     20 & $207.6$                      & $-2.375$      & $(1.362\lambda+7.829)\times 10^{-5}$\\
     30 & $2.661\times 10^3$           & $-3.625$      & $(8.487\lambda+70.02)\times 10^{-8}$\\
     40 & $3.337\times 10^4$           & $-4.875$      & $(5.468\lambda+58.78)\times 10^{-10}$\\
     50 & $4.142\times 10^5$           & $-6.125$      & $(3.577\lambda+47.40)\times 10^{-12}$\\
   \bottomrule
  \end{tabular}
 \end{table}

 Before doing so, we provide a comparison of precise numerical values for various physical quantities
 with their first order equivalents in Table~\ref{various_npoly}. One can see that the accuracy of the
 method does not depend strongly on the polytropic index and that relative errors are within a few percent
 for rings with a radius ratio of 0.9.
 \begin{table*}
  \centering
  \begin{minipage}{140mm}
  \caption{Physical quantities to the first order in $\sigma$ (bold faced type) are
           compared to the correct, numerically determined values (normal type) for
           given polytropic index $n$ and given radius ratio $\varrho_\text i/\varrho_\text o=0.9$.\label{various_npoly}}
  \begin{tabular}{ccccccc}\toprule
    $n$  & $\bar{M}$ & $\bar{\Omega}^2$ & $\bar{J}$ & $\bar{P}$ & $\bar{T}$
        & $\bar{W}$\\ \midrule
%
   0.5  & {\bf 103} & {\bf 0.0216}  & $\mathbf{3.14\times 10^3}$ & {\bf 59.2} & {\bf 231} & $\mathbf{-640}$ \\
   0.5  &      105 &       0.0213  &          $3.21\times 10^3$ &       59.9 &       235 &          $-650$\\[0.5ex]
   1    & {\bf 143} & {\bf 0.0151}  & $\mathbf{5.84\times 10^3}$ & {\bf 89.2} & {\bf 359} & $\mathbf{-986}$ \\
   1    &      144  &      0.0150   &         $5.95\times 10^3$  &      90.1  &      364  &         $-999$\\[0.5ex]
   3    & {\bf 356} & $\mathbf{4.56\times10^{-3}}$ & $\mathbf{3.53\times 10^4}$ & {\bf 264} & $\mathbf{1.19\times 10^3}$
                                     & $\mathbf{-3.17\times 10^3}$ \\
   3    &      359  & $4.52\times10^{-3}$          &         $3.58\times 10^4$  &      266  &         $1.20\times 10^3$
                                    &$\mathbf{-3.21\times 10^3}$ \\[0.5ex]
   5    & {\bf 714} & $\mathbf{1.58\times10^{-3}}$ & $\mathbf{1.44 \times10^3}$ & {\bf 570} & $\mathbf{2.86 \times 10^3}$
                                    & $\mathbf{-7.44\times 10^3}$ \\
   5    &      720  &         $1.56\times10^{-3}$  &         $1.46\times 10^3$  &      575  &          $2.89\times 10^3$
                                         &         $-7.51\times 10^3$ \\
    \bottomrule
  \end{tabular}
  \end{minipage}
 \end{table*}

 \section{The Limit of Infinite Polytropic Index}

\begin{figure}
  \centerline{\includegraphics{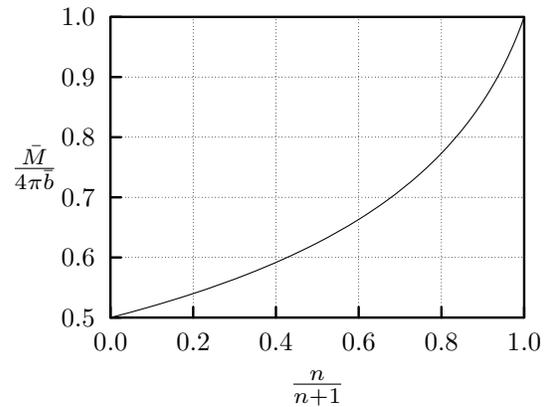}}
  \caption{The dimensionless mass divided by $4\pi\bar b$ in the thin ring limit is plotted versus $n/(n+1)$ over
           the whole range of polytropic indices $n\in [0,\infty)$. The points for $n=0$,
           $n=1$ and $n\to\infty$ are known analytically and the remainder of the curve was
           generated by solving the equation for $\tilde\mu_{00}$ numerically and making use
           of equation \eqref{mass_h00}.\label{M_4pic}}
 \end{figure}

 As $n$ tends to infinity, the polytropic equation \eqref{poly} shows us that pressure and
 density are proportional
 \begin{align}\label{isothermal} p=K\mu, \end{align}
 a case sometimes referred to as `isothermal' because such an equation holds for
 an ideal gas at constant temperature.
 Inserting this into equation \eqref{Laplace_Euler_polytrope} at leading
 order and again using the dimensionless coordinate $x$ yields
 \begin{align}
  4\pi x\mu_{00} + \frac{\d}{\d x}\left(x \frac{\d}{\d x}\ln\mu_{00} \right)=0.
 \end{align}
 The solution to this equation with our normalization $\mu_{00}(x=0)=1$ reads
 \begin{align}\label{mu_inf}
  \mu_{00} = \frac{4}{\pi^2\left(x^2+\frac{2}{\pi}\right)^2}.
 \end{align}
 The density and pressure fall to zero as $x\to\infty \Leftrightarrow r\to\infty$.
 Integrating over the density to calculate the normalized mass, one finds to leading order
 \begin{align}
  \bar M = 4\pi^2 \bar b \int_0^\infty \mu_{00} x\, \d x = 4\pi\bar b,
 \end{align}
 which can also be read off from equation \eqref{eq:P_M2} directly, by making use
 of $\bar M=\bar P$, which is self-evident upon taking \eqref{isothermal} into account.
 In Fig.~\ref{M_4pic}, the behaviour of $\bar M/4\pi\bar b$ can be followed from
 the homogeneous case, $n=0$, right up to the isothermal limit $n\to\infty$.

 Making use of \eqref{mu_inf}, we find that
 \begin{align}
   \lim_{n\to\infty} g=0,
 \end{align}
 where $g$ was defined in \eqref{def_g}. It thus follows from \eqref{g_minus_beta11} that
 \begin{align}
   \lim_{n\to\infty}\left( \beta_{11} +\frac{n}{8}\right)= \frac{1}{8}
 \end{align}
 as already suggested by the results of Table~\ref{beta11_Om2}. We can then see
 that the specific kinetic energy $\bar a^2\Omega_2$ tends to infinity such that
 for fixed $\lambda$
 \begin{align}
 \lim_{n\to\infty} \frac{2\pi\bar a^2\Omega_2}{n+4\lambda} = 1.
 \end{align}

 From the fact that $|\beta_{11}|$ tends to infinity, we can conclude that the range of $\sigma$ values
 for which the first order
 provides a good approximation shrinks to the point $\sigma=0$. This provides us with evidence
 suggesting that the deviation in a ring's cross-section from a circle becomes more
 pronounced at a given radius ratio as $n$ is increased. The value for $\varrho_\text i/\varrho_\text o$
 at which one reaches the mass-shedding limit presumably tends to 1 as $n$ tends to infinity.

 \section*{Acknowledgments}
   Many thanks to Professor R.\ Meinel for the helpful discussions.
   The authors are also grateful to Professor J.\ Ostriker for pointing
   out his work on this subject to us.
   Many of the computations in this paper made use of Maple\texttrademark. Maple is a
   trademark of Waterloo Maple Inc.
   This research was funded in part by the Deutsche Forschungsgemeinschaft
   (SFB/TR7--B1).

 \bibliographystyle{mn2e}
 \bibliography{Reflink} 

 \appendix
 \section{An Identity Relating Hypergeometric to Bessel Functions}\label{app:F}
 In order to prove the identity \eqref{F_identity}, we prove the more general
 identity
 \begin{align}\label{app:id}
  \begin{split}
   & \frac{3}{40}z\, _2\!F_3\left(\frac{5}{2},\frac{5}{2};3,3,\frac{7}{2};-z \right)
    - \frac{1}{2}\, _2\!F_3\left(\frac{3}{2},\frac{3}{2};2,2,\frac{5}{2};-z \right)\\
   &\qquad = \frac{\d}{\d z}\left[J_0(\sqrt z)^2 \right],
  \end{split}
 \end{align}
 for an arbitrary complex number $z$, from which \eqref{F_identity} follows immediately.

 We begin by using the differentiation properties of the hypergeometric functions, e.g.\
 7.2.3.47 in \citet{PBM90b}, to write
 \begin{align}
  \begin{split}
   & \frac{3}{40}\,\, _2\!F_3\left(\frac{5}{2},\frac{5}{2};3,3,\frac{7}{2};-z \right) z
    - \frac{1}{2}\,\, _2\!F_3\left(\frac{3}{2},\frac{3}{2};2,2,\frac{5}{2};-z \right)\\
   &\qquad = \frac{\d}{\d z}\Bigg[\,_2\!F_3\left(\frac{1}{2},\frac{1}{2};1,1,\frac{3}{2};-z \right)\\ &\qquad\qquad\qquad
                  - \frac{z}{3}\,\,_2\!F_3\left(\frac{3}{2},\frac{3}{2};2,2,\frac{5}{2};-z \right) \Bigg].
  \end{split}
 \end{align}
 With the integral identity 7.2.3.11, the term to be differentiated can
 be written as
 \begin{align}
  \begin{split}
   & _2\!F_3\left(\frac{1}{2},\frac{1}{2};1,1,\frac{3}{2};-z \right)
                  - \frac{z}{3}\,\,_2\!F_3\left(\frac{3}{2},\frac{3}{2};2,2,\frac{5}{2};-z \right)\\
   & \quad = \frac{1}{\pi}\int_0^1\!\!\!\int_0^1 \frac{J_0(2\sqrt{t_1 t_2 z})}{2\sqrt{t_1 t_2 (1-t_1)}}
            -\frac{z J_1(2\sqrt{t_1 t_2 z})}{2\sqrt{z(1-t_1)}} \,\d t_1 \, \d t_2,
  \end{split}
 \end{align}
 where we have made use of the identity (e.g.\ 7.13.1.1 in \citealt{PBM90b})
 \begin{align}
  _0\!F_1(b,-z) = \Gamma(b) z^{(1-b)/2} J_{b-1}(2\sqrt z).
 \end{align}
 The above double integral yields
 \begin{align}
  \begin{split}
   & \frac{1}{\pi}\int_0^1\!\!\!\int_0^1 \frac{J_0(2\sqrt{t_1 t_2 z})}{2\sqrt{t_1 t_2 (1-t_1)}}
            -\frac{z J_1(2\sqrt{t_1 t_2 z})}{2\sqrt{z(1-t_1)}} \,\d t_1 \, \d t_2\\
   & \qquad =\frac{1}{\pi}\int_0^1 \frac{J_0(2\sqrt{t_1 z})}{t_1 (1-t_1)} \d t_1 = J_0(\sqrt z)^2,
 \end{split}
 \end{align}
 thereby proving \eqref{app:id}.

 \label{lastpage}

\end{document}